\documentclass[twocolumn,showpacs,preprintnumbers,nofootinbib,superscriptaddress]{revtex4}
\usepackage{graphicx}
\usepackage{latexsym}
\usepackage{amssymb}
\usepackage{verbatim}
\def\lsim{\mathrel{\rlap{\lower4pt\hbox{\hskip1pt$\sim$}}
    \raise1pt\hbox{$<$}}}
\def\gsim{\mathrel{\rlap{\lower4pt\hbox{\hskip1pt$\sim$}}
    \raise1pt\hbox{$>$}}}
\def\sqr#1#2{{\vcenter{\vbox{\hrule height.#2pt
         \hbox{\vrule width.#2pt height#1pt \kern#1pt
         \vrule width.#2pt}
         \hrule height.#2pt}}}}

\def\beq{\begin{equation}}
\def\eeq{\end{equation}}
\def\beqa{\begin{eqnarray}}
\def\eeqa{\end{eqnarray}}
\begin{document}

\title{WMAP5 constraints on the unified model of dark energy and dark matter}

\author{T. Barreiro}
\altaffiliation[Also at Instituto de Plasmas e Fus\~ao Nuclear] {, Instituto Superior T\'ecnico, Lisboa.
 Email address: tiagobarreiro@fisica.ist.utl.pt}
 
 \affiliation{Dept. de Matem\'atica, Univ. Lus\'ofona de Humanidades e Tecnologias\\ Av. Campo Grande, 376, 1749-024 Lisboa.}

\author{O. Bertolami}
\altaffiliation[Also at Instituto de Plasmas e Fus\~ao Nuclear]{, Instituto Superior T\'ecnico, Lisboa. Email address:
orfeu@cosmos.ist.utl.pt}

\author{P. Torres}
\altaffiliation[Also at Centro de F\'isica Te\'orica e de Part\'iculas]{, Instituto Superior T\'ecnico, Lisboa. Email
address: torres@cftp.ist.utl.pt }

\affiliation{ Departamento de F\'isica, Instituto Superior T\'ecnico \\
Av. Rovisco Pais 1, 1049-001 Lisboa, Portugal}

\vskip 0.5cm

\date{\today}

\begin{abstract}

We derive constraints on the parameter space of the unified model of dark energy and dark matter, the Generalized Chaplygin Gas (GCG), from the amplitudes and positions of the first few peaks and first trough of the cosmic microwave background radiation (CMBR) power spectrum, using the latest WMAP five year data.

\vskip 0.5cm

\end{abstract}

 \pacs{98.80.-k,98.80.Cq,12.60.-i\hspace{4cm} Preprint DF/IST-1.2008}

\maketitle
\vskip 2pc
%%%%%%%%%%%%%%%%%%%%%%%%%%%%%%%%%%%%%%%%%%%%%%%%%%%%%%%%%%%%%%%%%%%%%%%%%%%
\section{Introduction}

The unified model of dark energy and dark matter \cite{GCG3,GCG3_2,bilic} is an interesting alternative to account the accelerated expansion of the universe. It is shown to be consistent with first year WMAP and other CMBR data \cite{GCG6,GCG_and_WMAP,amendola}, gravitational lensing \cite{GCG7}, supernova data \cite{realGCG,GCG2}, cosmic topology \cite{GCG4}, gama ray bursts \cite{GCG5} and structure formation (see discussions in Refs. \cite{sandvik,GCG1}). The compatibility of the GCG scenario with the neutrino background radiation has been studied in Refs. \cite{bernardini1,bernardini2}.

The GCG behavior is described by an exotic perfect fluid with the following equation of state \cite{GCG3,GCG3_2} 
\beq 
p_{ch} = -\frac{A}{\rho ^{\alpha}_{ch}}~, 
\label{GCGes} 
\eeq 
\noindent 
where $A$ is a positive constant and $\alpha$ is a constant in the range $0\leq \alpha \leq 1$.
The covariant conservation of the energy-momentum tensor for an homogeneous and isotropic
spacetime implies, in terms of the scale factor, $a$, that:
\beq 
\rho_{ch} = \Bigl[ A + \frac{B}{a^{3(1+\alpha)}}\Bigr]^{\frac{1}{1+\alpha}}~,
\label{GCG_dens1}
\eeq
\noindent where $B$ is an integration constant. That is, at early times the GCG behaves like non-relativistic matter, while at late times it mimics a cosmological constant. This behavior is maintained even after considering energy density perturbations \cite{GCG3_2,bilic}.

In this work, we update the analysis carried out in Refs. \cite{GCG_and_WMAP,GCG6} constraining the parameters of the GCG model. We use the bounds from the positions of the peaks and troughs of the CMBR power spectrum that have an increased precision from the WMAP Three Year Observations
and we add the bounds arising from the baryon acoustic oscillations (BAO). In this scenario we consider the GCG to be a unified dark matter/dark energy fluid with no additional cold dark matter.

We also perform a separate analysis using the full CMBR spectrum. In this case, we include an additional cold dark matter fluid, and allow for the full perturbation of the GCG fluid, following Refs. \cite{amendola,perturbchap}. We use a modified CAMB code \cite{camb} to compute the theoretical CMBR spectrum and the recently released WMAP5 Five Year Observations \cite{WMAP5} to obtain the likelihoods. We then perform a Markov chain Monte Carlo analysis with the COSMOMC code \cite{cosmomc}, using a Metropolis algorithm, to constrain the GCG parameters.

In section \ref{theory} we describe the theoretical framework of our study; our results are presented in section \ref{results}. Finally, our conclusions are discussed in section \ref{conclusion}.

%%%%%%%%%%%%%%%%%%%%%%%%%%%%%%%%%%%%%%%%%%%%%%%%%%%%%%%%%%%%%%%%%%%
\section{CMBR constraints on the GCG model}\label{theory}

\subsection{Constraints from peak and trough locations}\label{peakconstraint}

The CMBR peaks arise from acoustic oscillations of the primeval plasma just before the universe becomes transparent. The scale of these oscillations is set by the acoustic scale, $l_A$, which, for a flat universe, is given by

\beq
l_A=\pi \frac{\tau_0 - \tau_{ls}}{\overline{c}_S \tau_{ls}}~,
\label{lA1}
\eeq
\noindent where $\tau = \tau_{ls}^{-1} \int a^ {-1}dt$ is the conformal time, $\tau_0$ and $\tau_{ls}$ being its value at present and at last scattering respectively, while $\overline{c}_S$ is the average sound speed before decoupling:
\beq
\overline{c}_S \equiv \tau_{ls}^{-1}\int_0^{\tau_{ls}} c_S d\tau~,
\label{c_S_average}
\eeq
\noindent where
\beq
c_S^{-2} = 3 + {9\over4} \frac{\rho_b(t)}{\rho_\gamma(t)}~,
\label{c_S}
\eeq
\noindent with $\rho_b/\rho_{\gamma}$ being the ratio of baryon to photon energy density.

In an idealized model of the primeval plasma, there is a straightforward relationship between the location of the $m$-th peak and the acoustic scale, namely $l_m \approx m l_A$. However, the peak positions are shifted by several effects which can be estimated by parametrizing the location of the $m$-th peak, $l_m$, as in Refs. \cite{Doran1,Hu,Doran3}
\beq
l_{pm} \equiv l_A (m - \varphi_m) \equiv l_A (m - \overline{\varphi} - \delta \varphi_m)~,
\label{l_pm}
\eeq

\noindent where $\overline{\varphi} \equiv \varphi_1$ is the overall peak shift and $\delta\varphi_m \equiv \varphi_m -\overline{\varphi}$ is the relative shift of the $m$-th peak relative to the first one. Eq.~(\ref{l_pm}) can also be used for the position of troughs by setting $m = 3/2$ for the first one. We use the fitting formulae of Refs. \cite{Doran1,Hu,Doran3,reese02} for the shifts and the theoretical estimate of the peak locations as described in the Appendix.

The energy density, Eq. (\ref{GCG_dens1}), can be rewritten as
\beq
\rho_{ch}=\rho_{ch0} \Bigl(A_S + \frac{1-A_S}{a^{3(1+\alpha)}} \Bigr)^{1\over{1+\alpha}}~,
\label{GCG_dens2}
\eeq

\noindent through the definitions $A_S \equiv A/\rho_{ch0}^{1+\alpha}$ and $\rho_{ch0}=(A+B)^{1\over{1+\alpha}}$. It follows for the expansion rate
\beq
H^2 = \frac{8\pi G}{3} \Bigl[ \frac{\rho_{r0}}{a^4} + \frac{\rho_{b0}}{a^3} + \rho_{ch0} \Bigl(A_S + \frac{1-A_S}{a^{3(1+\alpha)}} \Bigr)^{1\over{1+\alpha}}\Bigr]~,
\label{friedmann1}
\eeq
where $\rho_{b0}$ and $\rho_{r0}$ are the baryon and radiation energy densities at present. We do not include cold dark matter in this model, its role being taken by the GCG.

It is worth mentioning that $0\leq A_S\leq 1$ and that, for $A_S=0$, the GCG behaves as dust while for $A_S=1$ it behaves like a cosmological constant. The GCG model matches the $\Lambda CDM$ model for $\alpha=0$ and $A_S=1$.

Using the fact that $\rho_{r0}/\rho_{ch0} = \Omega_{r0}/(1 - \Omega_{r0} - \Omega_{b0})$ and $\rho_{b0}/\rho_{ch0} = \Omega_{b0}/(1 - \Omega_{r0} - \Omega_{b0})$, 

\beq
H^2=\Omega_{ch0}H_0^2 a^{-4} X^2(a)~,
\label{H_X}
\eeq
\noindent where

\beqa
X^2(a)=\frac{\Omega_{r0}}{1 - \Omega_{r0} - \Omega_{b0}} + \frac{\Omega_{b0} a}{1 - \Omega_{r0} - \Omega_{b0}} \nonumber\\
+ a^4\Bigl(A_S + \frac{1-A_S}{a^{3(1+\alpha}}\Bigr)^{\frac{1}{1+\alpha}}~.
\label{X-a}
\eeqa

\noindent Moreover, since $H^2 = a^{-4}\left( \frac{da}{d\tau}\right)^2$, we get

\beq
d\tau = \frac{da}{\Omega_{ch0}^{1/2} H_0 X(a)}~,
\label{d_tau}
\eeq
\noindent so that

\beq
l_A = \frac{\pi}{\overline{c}_S} \Bigl[ \int_0^1\frac{da}{X(a)} \Bigl( \int_0^{a_ls} \frac{da}{X(a)} \Bigr)^{-1} - 1\Bigr]~,
\label{lA2}
\eeq

\noindent where $a_{ls}$ is the scale factor at last scattering, for which we use the fitting formulae
\cite{Doran1,Hu,Doran3,reese02}:

\beq
a_{ls}^{-1} = z_{ls} = 1048\Bigl[1 + 0.00124 \omega_b^{-0.738}\Bigr]\Bigl[1 + g_1 \omega_m^{g2}\Bigr]~,
\label{fit_als}
\eeq
\noindent where
\beqa
g_1 &=& 0.0783 \omega_b^{-0.238}\Bigl[1 + 39.5 \omega_b^{0.763}\Bigr]^{-1}~,\nonumber\\
g_2 &=& 0.56 \Bigl[1 +1 21.1 \omega_b^{1.81}\Bigr]^{-1}~,
\eeqa
\noindent and $\omega_{b,m} \equiv \Omega_{b,m} h^2$.

For the position of the peaks we consider the three year WMAP measurements, which show a considerable precision on the locations of the first two peaks and the first trough, namely \cite{WMAP3}:

\beqa
l_{p1} &=& 220.8 \pm 0.7~,\nonumber\\
l_{p2} &=& 530.9 \pm 3.8~,\nonumber\\
l_{d1} &=& 412.4 \pm 1.9~,
\label{2peaks}
\eeqa

\noindent at $1\sigma$.

We also consider the bound from the baryon acoustic peak \cite{baryonpeak}. Its position is related to the quantity
\beq
{\cal A}=\sqrt{\Omega_m}\left(\frac{H_0}{H(z_{lrg})}\right)^{1/3}\left[\frac{1}{z_{lrg}}\int_0^{z_{lrg}} \frac{H_0}{H(z')}dz')\right]^{2/3}~,
\label{baryoneq}
\eeq
which takes the value
\beq
A_0 = 0.469 \pm 0.017~.
\label{baryonbounds}
\eeq
\noindent where $z_{lrg} = 0.35$ is the redshift from the Sloan Digital Sky Survey for luminous red galaxies.

Combining Eqs. (\ref{l_pm}) and (\ref{lA2}) and the fitting formulae of Refs  \cite{Doran1,Hu,Doran3,reese02}, shown in the Appendix, we search for the combination of GCG model parameters that is consistent with the observational bounds, taking also into account the theoretical error estimates given in the Appendix. For this analysis we fixed the value of the spectral index $n_s = 0.963$ to the WMAP mean, and we used $\omega_{r0} = 4.1532 \times 10^{-5}$ and $\omega_{b0} = 0.027$ for the present energy densities of radiation and baryons respectively.

Our results are shown in section \ref{results}.

%%%%%%%%%%%%%%%%%%%%%%%%%

\subsection{Likelihood analysis}

Besides the peaks and trough analysis described above we also consider
the full temperature and polarization spectrum to constrain the GCG model with the latest CMBR data from WMAP5.
We relax our value for the cold dark matter, so that we are now really constraining a mixed model, where the GCG will be mainly responsible for the dark energy. In order to do this, we obtain the theoretical CMBR spectrum using a modified CAMB \cite{camb} code to include the GCG evolution in the background, Eqs. (\ref{GCG_dens2})-(\ref{friedmann1}), as well as its perturbations \cite{perturbchap}. Using derivatives with respect to conformal time and defining the conformal time Hubble parameter, ${\cal H} = a' / a$, (where the primes denote differentiation with respect to the conformal time) we have

\beqa
{\delta}' &=& -(1+w)\left( \theta+\frac{\dot{h}}{2} \right) - 3 (c_s^2  - w) {\cal H} \delta~,\\
{\theta}' &=& (3 c_s^2 - 1) {\cal H} \theta + \frac{c_s^2}{(1-w)}k^2 \delta~,
\label{GCGperturb}
\eeqa
where $\delta$ and $\theta$ are the density and velocity divergence perturbations for the GCG fluid, and
\beqa
w & = & \frac{p_{ch}}{\rho_{ch}}~,\\
c_s^2 & = & - w \alpha
\eeqa
are, respectively, the GCG fluid  equation of state and  speed of sound. We assume that the shear of the GCG fluid vanishes.

We included these equations into the CAMB code \cite{camb}, and used the latest WMAP five year data likelihood code \cite{WMAP5} to compute the likelihoods for this model. We used the COSMOMC \cite{cosmomc} code to perform a Markov chain Monte Carlo analysis on our parameter space. For simplicity we fixed the inflationary input (namely, the spectral index $n_s = 0.963$), and allowed changes in the parameters $\alpha$ and $A_S$ for the GCG fluid, $\omega_b$ and $\omega_c$ for the present baryon and cold dark matter energy density, $\tau$ the optical depth and $h$ the Hubble constant. We fixed the radiation value at the present to be $\omega_{r0} = 4.1534 \times 10^{-5}$.

We present our results in the next section.

%%%%%%%%%%%%%%%%%%%%%%%%%%%%%%%%%%%%%%%%%%%%%%%%%%%%%%%%%%%%%%%%%%
%%%%%%%%%%%%%%%%%%%%%%%%%%%%%%%%%%%%%%%%%%%%%%%%%%%%%%%%%%%%%%%%%%
\section{Results and Discussion}\label{results}

\subsection{Peak and trough constraints}\label{picos}

%%%%%%%%%%%%%%%%%%%%%%%%%%%%%%%%%%%%%%%%%%%%%%%%%%%%%%%%%%%%%%%%%%
\begin{figure}[h]
\includegraphics[width=7cm]{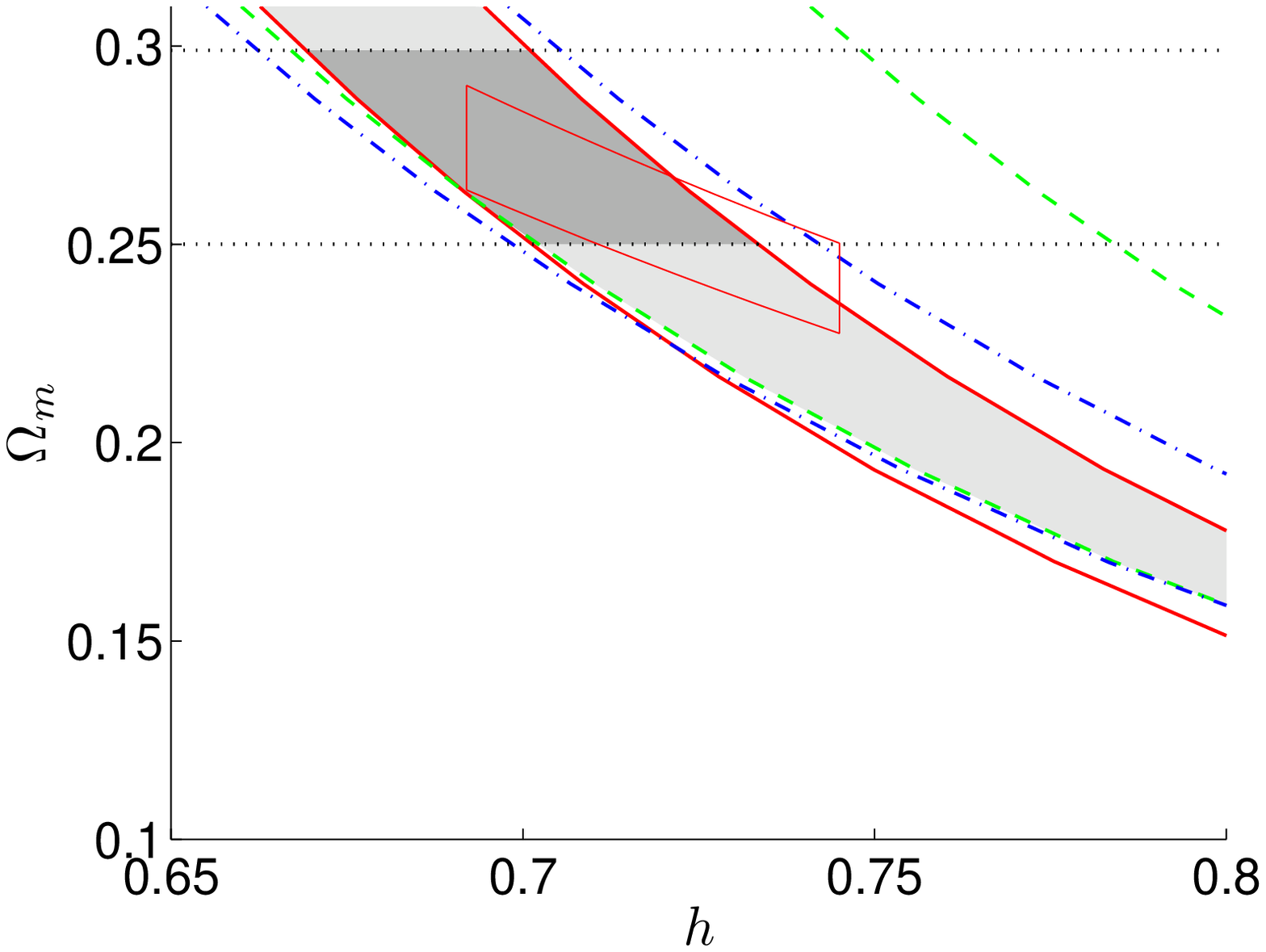}
\includegraphics[width=7cm]{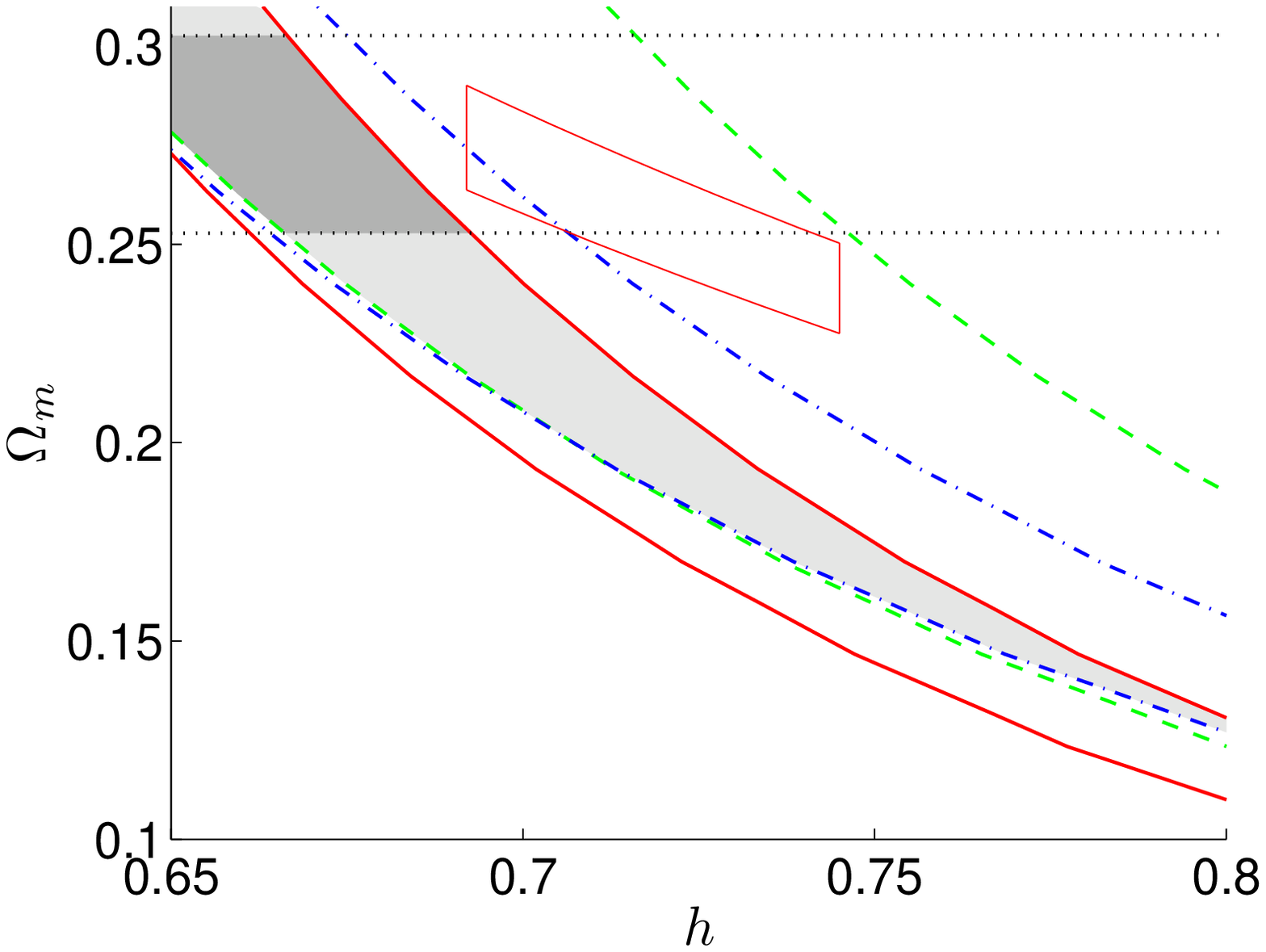}
\caption[fig:omega_vs_h]{Contour plots of the first two Doppler peaks and first trough locations, as well as for the baryon acoustic peak, in the ($\Omega_m$,$h$) plane for the GCG model, with $n_S=0.963$ for $\alpha=0$ (top graphic) and $\alpha=0.2$ (bottom figure). Full, dashed, dot-dashed and dotted contours correspond to observational bounds on $l_{p1}$, $l_{p2}$, $l_{d1}$ (Eqs. (\ref{2peaks})) and the baryon acoustic peak (Eq. (\ref{baryonbounds})), respectively. The box indicates the bounds on $h$ and $\Omega_m h^2$ from the WMAP five year data. Darker and lighter shaded areas correspond to the allowed regions with and without BAO, respectively.}
 \label{fig:omega_vs_h}
\end{figure}

Fig. \ref{fig:omega_vs_h} shows the contour plots of the two first Doppler peaks and the first trough locations, as well as the baryon acoustic oscillations (BAO) peak, in the $(\Omega_m,h)$ plane for the GCG model. Since we do not have cold dark matter in this model, the $\Omega_m$ parameter is really an estimate of the ``matter'' component of the GCG with the baryon energy density, namely through the relation to $A_S$ given by
\begin{equation}
\Omega_m = 1 - \Omega_r - A_S (1 - \Omega_b - \Omega_r)~.
\end{equation}

We see that the dependance of the BAO contour on $h$ is very slight, so the BAO constraint works as a lower bound on the value of $\Omega_m$. The boxes depicted in the figures are the \mbox{1-sigma} bounds on $h$ and $\Omega_m h^2$ from the WMAP five year data fit to a $\Lambda$CDM model, as indicated in the caption.
We can see that for $\alpha=0$ (top figure) the results are compatible with the observations, as expected; notice also that the second peak is not measured with a sufficient accuracy to further constrain the  results from the other bounds.

For $\alpha=0.2$ (bottom figure), we see that the BAO bound combined with the first peak bound pushes  $h$ to smaller values, making the model less viable. Higher values of $\alpha$ are almost ruled out from the observations (see below).

%%%%%%%%%%%%%%%

\begin{figure}[h]
\includegraphics[width=7cm]{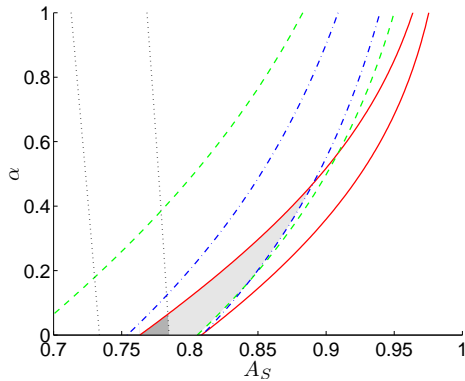}
\caption[fig:alfa_vs_as]{Contour plots of the first two Doppler peaks and first trough locations, as well as for the baryon acoustic peak, in the ($A_S$,$\alpha$) plane for the GCG model, with $n_S=0.963$ and $h=0.719$. Full, dashed, dot-dashed and dotted contours correspond to observational bounds on, respectively, $l_{p1}$, $l_{p2}$, $l_{d1}$ (Eqs. (\ref{2peaks})) and the baryon acoustic peak (Eq. (\ref{baryonbounds})). Darker and lighter shaded areas correspond to the allowed regions with and without BAO, respectively.}
 \label{fig:alfa_vs_as}
\end{figure}

Fig. \ref{fig:alfa_vs_as} shows the same bounds in the $(A_S,\alpha)$ plane. We now fix  $h=0.719$, still using $n_s = 0.963$. We see that the first peak and trough constraints allow large $A_S$, $A_S \simeq 0.9$, and $\alpha \lsim 0.4$ (again, the second peak location does not noticeably change our results). Introducing 
the BAO constraint, however, severily affects these results, requiring $\alpha \lsim 0.1$ and $As \simeq 0.77$.

%Note that these results were obtained considering a unified dark energy/dark matter scenario for the GCG. We assume that the usual troubles associated with this scenario \cite{sandvik} are avoided in some way, see e.g. \cite{revival}.

In comparing these results with \cite{GCG_and_WMAP}, we have to consider that we are now including the estimated theoretical errors for the fitting formulae \cite{Doran1}. Taking this into account shows that the new WMAP results make a considerable improvement in constraining the parameter space of the model. The BAO bounds add a significant contribution to these constraints, further reducing the allowed region of parameter values.

%%%%%%%%%%%%

\subsection{Likelihood analysis}

We start recalling that in this approach, we use the full set of perturbation equations for the GCG, so we are mainly considering the GCG fluid to act as the dark energy component. That is, we allow a separate cold dark matter energy density as a free parameter in our model.

\begin{figure}[h]
\includegraphics[width=7cm]{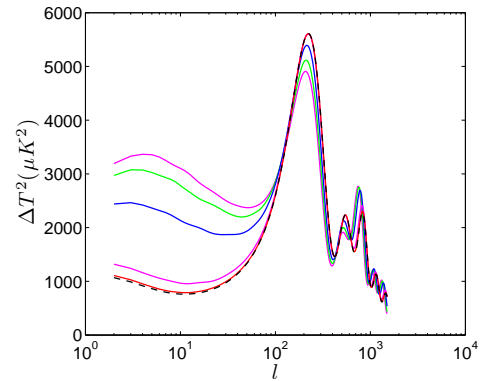}
\caption[fig:spectra_as]{The spectrum for the GCG model, compared with a $\Lambda CDM$ model (dashed curve), for $\alpha=0.5$, $n_S=0.963$, $\tau=0.087$ and $h=0.719$, for $A_S=0.7,~0.8,~0.9,~0.99,~0.999$, from top to bottom, respectively (at the plateau).}
 \label{fig:spectra_as}
\end{figure}

\begin{figure}[h]
\includegraphics[width=7cm]{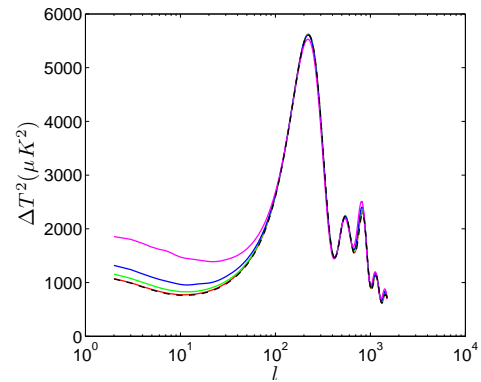}
\caption[fig:spectra_alpha]{The spectrum for the GCG model, compared with a $\Lambda CDM$ model (dashed curve), for $A_S=0.99$, $n_S=0.963$, $\tau=0.087$ and $h=0.719$, for $\alpha=1,~0.5,~0.2,~0.1,~0$, from top to bottom, respectively (at the plateau).}
 \label{fig:spectra_alpha}
\end{figure}

In Figs. \ref{fig:spectra_as} and \ref{fig:spectra_alpha} we show the WMAP normalized power spectrum computed with CAMB . We used the WMAP five year mean values for the spectral index, $n_S=0.963$, the optical depth to the last scattering surface, $\tau=0.087$, and the Hubble constant, $h=0.719$. We can see that the dependance of the GCG  model on $A_S$ is stronger than on $\alpha$. The first and second peak move to the left as we decrease $A_S$ or as we increase $\alpha$, in consistency with the results from the previous sub-section. For the same parameter changes, the third peak moves to the right.
The ratio between the first peak amplitude and the plateau gets smaller with decreasing $A_S$ or increasing $\alpha$, due to the integrated Sachs-Wolfe effect (ISW) \cite{amendola,perturbchap}. 
The first and second peak intensities increase for greater values of $A_S$ and smaller values of $\alpha$, while the plateau and third peak decrease.
We can see that the model favours $A_S$ values close to one and small $\alpha$ values.

%%%%%%%%%%%%%%%

%%%%%%%%%%%%%%%

\begin{figure}[t]
\includegraphics[width=7cm]{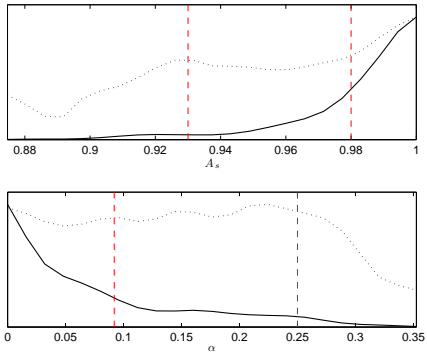}
\caption[fig:likelihood01]{Marginalized posterior probability densities for the model parameters. Dotted lines are the likelihood functions. The vertical dashed lines mark the $1\sigma$ and $2\sigma$ confidence levels.}
 \label{fig:likelihood01}
\end{figure}

Fig. \ref{fig:likelihood01} shows the marginalized posterior probability densities for $\alpha$ and $A_S$. We find the bounds
\beqa
\alpha &<& 0.25~,\nonumber\\
0.93 &<& A_S < 1~,
\label{limits95}
\eeqa
\noindent at $2\sigma$.

The obtained mean values for the other parameters are $\omega_c =0.0935$, $\omega_b =0.0225$, $\tau=0.70$ and $h=0.74$.

The difference in the bounds of $A_S$ obtained by the previous approach and this one is easily understood. In the first case we do not consider the dark matter component, so the GCG is modeling both, dark matter and dark energy. In the second approach, dark matter is included, so the GCG contribution to this component is not the dominating one. This tends to produce a higher lower bound for $A_S$.

%%%%%%%%%%%%%%%

%%%%%%%%%%%%%%%

%%%%%%%%%%%%%%%%%%%%%%%%%%%%%%%%%%%%%%%%%%%%%%%%%%%%%%%%%%%%%%%%%%
\section{Conclusions}\label{conclusion}

In the present work we have considered the position of the first two peaks and trough from WMAP three year data and other CMBR experiments that best fit WMAP five year data for $h$ and $n_S$. 
Following the analysis of Refs. \cite{GCG6,GCG_and_WMAP}, the new WMAP CMBR data shows that the GCG model is compatible with observations for $\alpha \lsim 0.4$. The inclusion of the BAO bound further constraints the model, the allowed region of the GCG parameters being $\alpha \lsim 0.1$ and $A_S \simeq 0.77$.

Secondly, we considered the full CMBR spectrum allowing the full perturbation of the GCG fluid. We used a modified CAMB code to compute the theoretical spectrum and the COSMOMC code to perform a Markov chain Monte Carlo analysis with a Metropolis algorithm. The bound on the GCG parameters obtained were $\alpha < 0.25$ and $A_S > 0.93$, at $2\sigma$.

%%%%%%%%%%%%%%%%%%%%%%%%%%%%%%%%%%%%%%%%%%%%%%%%%%%%%%%%%%%%%%%%%%%%%%%
\begin{acknowledgments}
The work of T. Barreiro was partially financed by the  Funda\c c\~ao para a 
Ci\^encia e a Tecnologia (FCT, Portugal) under the grant PPD/3512/2000.
The work of O.B. is partially supported by the FCT project POCI/FIS/56093/2004.
The work of P.T. is fully supported by FCT under the grant SFRH/BD/25592/2005.

\end{acknowledgments}
%%%%%%%%%%%%%%%%%%%%%%%%%%%%%%%%%%%%%%%%%%%%%%%%%%%%%%%%%%%%%%%%%%
\appendix*
\section{}

We have used the analytic approximations for the phase shifts found in
 Refs.~\cite{Doran2,Hu1996}, which we reproduce here for convenience.
 The overall phase shift is given by
\beq
{\bar\varphi}=(1.466-0.466 n_s)\Bigl[a_1 r_*^{a_2}
+ 0.291 {\bar\Omega}_{\phi}^{ls} \Bigr]~,
\label{eq:phibar}
\eeq   
where

\beqa
a_1 &=& 0.286+0.626\omega_b\nonumber\\
a_2 &=& 0.1786-6.308\omega_b+174.9\omega_b^2-1168\omega_b^3
\label{eq:as}
\eeqa
are fitting coefficients, $\omega_b=\Omega_b h^2$,

\beq
{\bar \Omega}_{\phi}^{ls}=\tau_{ls}^{-1} \int_0^{\tau_{ls}} \Omega_\phi(\tau)
 d\tau~,
\label{eq:baro}
\eeq
and

\beq
r_*\equiv \rho_{rad}(z_{ls})/\rho_{m}(z_{ls})
\eeq
is the ratio of radiation to matter  at decoupling. A suitable fitting formula for $z_{ls}$
 is \cite{Hu1996}

\beq
z_{ls}=1048 [1+0.00124 w_b^{-0.738}] [1+g_1 w_m^{g_2} ]~,
\label{eq:zls}
\eeq
where

\beqa
g_1 &=& 0.0783 w_b^{-0.238} [1+39.5 w_b^{0.763}]^{-1}~,\nonumber\\
g_2 &=& 0.56 [1+21.1 w_b^{1.81}]^{-1}~.
\label{eq:coef}
\eeqa

The relative shift of the first acoustic peak is zero, $\delta\varphi_1=0$,
and the relative shifts of the  second and third peaks are given by

\beq
\delta\varphi_2=c_0-c_1r_*-c_2r_*^{-c_3}+0.05(n_s-1)~,
\label{eq:delphi2}
\eeq
with
\beqa
c_0 &=& -0.1+\left(0.213-0.123{\bar\Omega}_{ls}^\phi \right)\nonumber\\
        &&\times\exp\left\{-\left(52-63.6 {\bar\Omega}_{ls}^\phi \right)
          \omega_b\right\},\nonumber\\
c_1 &=& 0.015+0.063\exp\left(-3500\omega_b^2\right), \nonumber\\
c_2 &=& 6\times 10^{-6}+0.137(\omega_b-0.07)^2,\nonumber\\
c_3 &=& 0.8+ 2.3 {\bar\Omega}_{ls}^\phi
+\left(70-126{\bar\Omega}_{ls}^\phi\right)\omega_b~,
\label{eq:cs}
\eeqa
and 

\beq
\delta\varphi_3=10-d_1r_*^{d_2}+0.08(n_s-1)~,
\eeq
with
\beqa
d_1 &=& 9.97+\left(3.3-3{\bar\Omega}_{ls}^\phi\right)\omega_b,\nonumber\\
d_2 &=& 0.0016-0.0067{\bar\Omega}_{ls}^\phi +
\left(0.196-0.22{\bar\Omega}_{ls}^\phi\right)\omega_b\nonumber\\
&& + \left(2.25+2.77{\bar\Omega}_{ls}^\phi\right)\times  10^{-5}\omega_b^{-1}.
\label{eq:ds}
\eeqa

The relative shift of the first trough is given by

\beq
\delta\varphi_{3/2}=b_0+b_1r_*^{1/3}\exp(b_2r_*)+0.158(n_s-1)~
\label{eq:delphi}
\eeq
with 

\beqa
b_0 &=& -0.086-0.079 {\bar\Omega}_{\phi}^{ls}
       -\left( 2.22-18.1{\bar\Omega}^{ls}_\phi \right) \omega_b\nonumber\\
    &&-\left(140+403{\bar\Omega}^{ls}_\phi \right)\omega_b^2~,\nonumber\\
b_1 &=& 0.39-0.98{\bar\Omega}^{ls}_\phi -\left(18.1-29.2{\bar\Omega}_{ls}^\phi\right)\omega_b\nonumber\\
    && +440\omega_b^2~,\nonumber\\
b_2 &=& -0.57-3.8\exp({-2365\omega_b^2})~.
\label{eq:bs}
\eeqa

The estimate for the accuracy of the fitting formulae is given at $1\sigma$ by \cite{Doran1}:

\beqa
&&\Delta\overline\phi = 0.0031~,\nonumber\\
&&\Delta\delta\phi_2 = 0.0044~,\nonumber\\
&&\Delta\delta\phi_{3/2} = 0.0039~.
\label{phi_error}
\eeqa

%%%%%%%%%%%%%%%%%%%%%%%%%%%%%%%%%%%%%%%%%%%%%%%

\end{document}